\documentclass{aa}  

\usepackage{graphicx}
\usepackage{txfonts}
\usepackage{hyperref}
\usepackage{color}

\newcommand{\pr}{$^{\prime}$}

\newcommand{\hii}{H~{\sc ii} }

\begin{document}

\title{Resolved properties of a luminous "hinge clump" in the compact group of galaxies NGC\,6845}


\author{Daniela E. Olave-Rojas\inst{1}
        \and
        José A. Hernandez-Jimenez \inst{2}
        \and
        Sergio Torres-Flores\inst{3}
        \and
        Marcelo D. Mora\inst{4}
        \and
        Veronica Firpo\inst{5}
          }
\institute{Departamento de Tecnologías Industriales, Facultad de Ingeniería, Universidad de Talca, Los Niches km 1, Curicó, Chile\\
              \email{daniela.olave@utalca.cl}
         \and
         Universidade do Vale do Paraíba, Instituto de Pesquisa e Desenvolvimento, Avenida Shishima Hifumi, 2911, São José dos Campos, SP 12244-000, Brazil
         \and
         Departamento de Astronomía, Universidad de La Serena, Avda. Juan Cisternas 1200, La Serena, Chile
         \and
         Las Campanas Observatory, Carnegie Institution for Science, Colina El Pino S/N, La Serena 1700000, Chile
         \and
         Gemini Observatory/NSF’s NOIRLab, Casilla 603, La Serena, Chile
             }

\date{Received September 15, 1996; accepted March 16, 1997}


 
\abstract
   {Compact groups of galaxies are unique places where galaxy-galaxy interactions play a mayor role on the evolution of its members. These strong gravitational encounters can induce star formation bursts.}
   {We study the properties of one of the most luminous "hinge clumps", located on the compact group of galaxies NGC\,6845.}
   {Using integral field spectroscopy from GMOS/Gemini, complemented with archival MUSE data,  we obtain oxygen abundances, ages, star formation rates, velocity fields and we also performed a single stellar populations modeling to understand the star formation history of the "hinge clump" localized in NGC\,6845.}
   { We found that the "hinge clump" sits in a tail, having a star formation rate of 3.4 M$_{\odot}$ yr$^{-1}$, which is comparable with a few other extreme cases, e.g., the star clusters in the Antennae galaxy and other reported "hinge clumps" in the literature. In fact, this clump represents $\sim$15\% of total SFR of NGC\,6845A. Large-scale modeling of the observed velocity field of NGC\,6845A rules out the scenario on which this "hinge clump" was a satellite galaxy. Indeed, its kinematics is compatible with the galactic disk of NGC\,6845A. Its abundance, mean value of 0.4Z$\odot$, is also consistent with the metallicity gradient of the galaxy.}
    {Our analysis, suggest that the hinge clump is formed by multiple stellar populations instead of a single burst, thus having a large range of ages. We found that central clump is encompassed by a ring-like structure, suggesting that the ring-like structure represents a second-generation of star formation. In addition, the analysis of the diagnostic diagram indicates that this central region can also be being ionized by shock from stellar and supernovae winds.  Finally, the derived SFR density $\Sigma$ = 9.7 M$_{\odot}$ yr$^{-1}$ kpc$^{-2}$ of the central clump, place it in starburst regime, where gas inflows should provide gas to maintain the star formation. This work shows a resolved example of an extreme localized starburst in a compact group of galaxies.}


\maketitle
%

\section{Introduction}

It is well known that interacting galaxies provide an extraordinary place to study galaxy transformation and evolution. In this sense, different physical phenomena occurs when galaxies interact. For instance, interactions can produce violent central starbursts, which significantly enhance the integrated star formation in galaxies (e.g. \citealt{2002Bekki}, \citealt{2016Renaud}, \citealt{2022Shah}). In the same context, an important fraction of this enhanced star formation can be linked to the formation of massive compact star clusters (e.g. \citealt{2010Whitmore}, \citealt{2021Linden}, \citealt{2022He}). These interactions also produce tidal tails, where a low-level star formation can be detected (e.g. \citealt{Knierman03}, \citealt{Mullan11}), showing that interacting galaxies provided us a natural place to study star formation at different levels.  

\begin{figure*}
\includegraphics[width=0.95\textwidth]{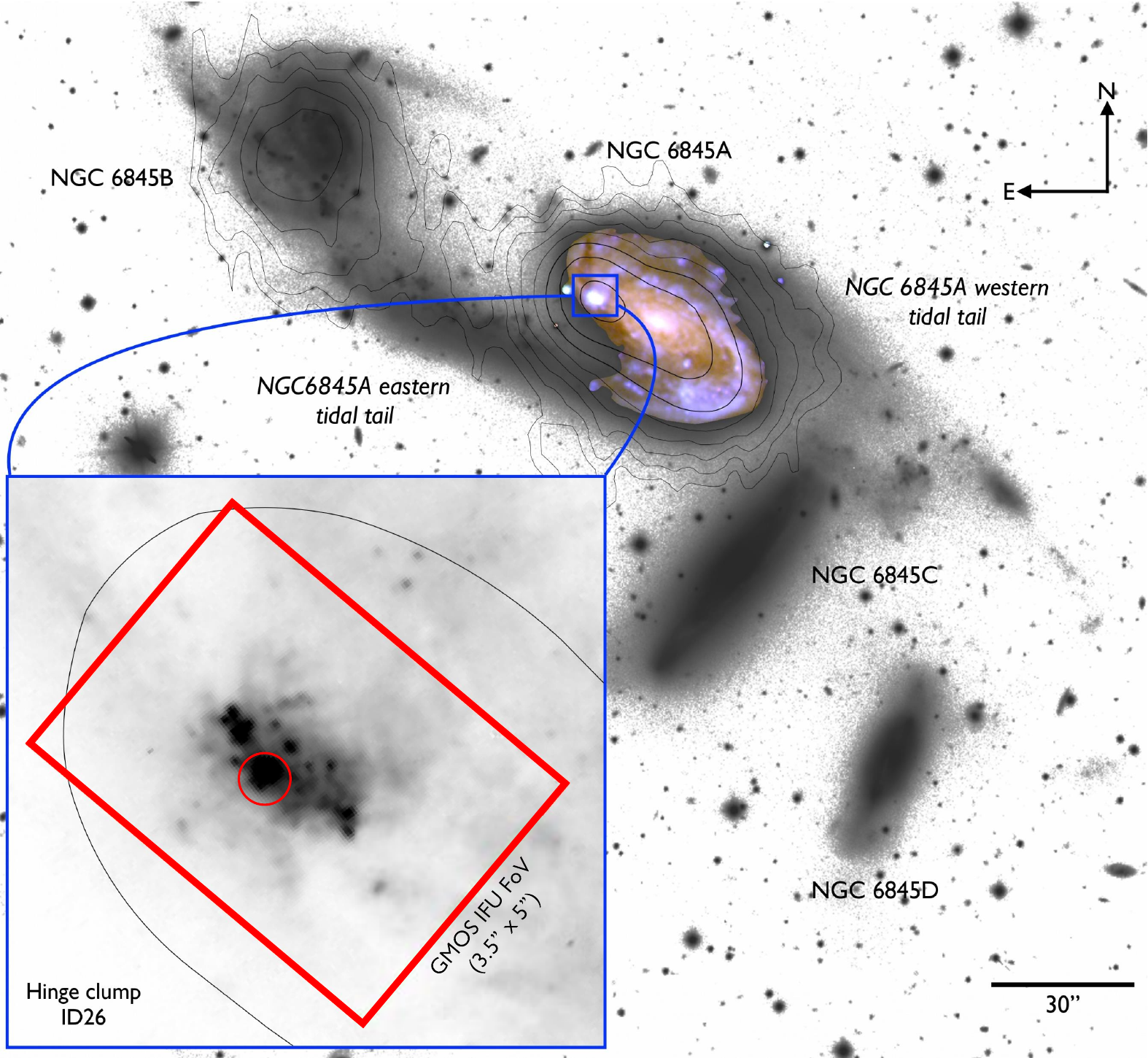}
\caption{Gemini/GMOS \textit{r'}-band high-contrast image of the compact group of galaxies NGC\,6845. The central region of NGC\,6845A is overplotted with an \textit{u\pr}, \textit{g\pr} and \textit{r\pr}-band composite image. The extra-nuclear star-forming region (hinge clump ID26) is identified in a blue box. Optical \textit{HST} archival imaging (\textit{F606W}) highlights the central structure of this star-forming complex (zoomed blue box), where we have over-plotted the Gemini/GMOS IFU field of view. Black contours represent the 20cm continuum taken from \citet{2003Gordon}. Red circle in the zoomed box corresponds to the peak of the continuum $H\alpha$ emission (cf, Fig. \ref{fig_line_fluxes}).}
\label{figfov}
\end{figure*}

In the case of tidal tails, different authors have found from tidal dwarf galaxies (TDGs, e.g. \citealt{Duc98}, \citealt{MdeO06}, \citealt{Bournaud07}) to low-mass star-forming regions and stellar clusters (e.g. \citealt{2004Li}. \citealt{2007Trancho}, \citealt{2008Bournaud}). Most of these systems display quite low star formation rates. In the case of TDGs, which are typically formed at the tip of tidal tails, some authors argue that these systems does not contribute significantly to the population of dwarf galaxies (e.g \citealt{2017Ploeckinger}), while other authors suggest that these systems will fade out in a couple of Gyrs (e.g. \citealt{2021Roman}). However, at the base of tidal tails the scenario is quite different. \cite{Hancock09} reported two luminous star-forming regions at the base of the northern and southern tidal tails of NGC\,4017. They named these sources as ``hinge clumps'', which have H$\alpha$ luminosities greater than 10$^{40}$ erg s$^{-1}$. Therefore, these sources appears as extra-nuclear starburst regions. More recently, \citep{2014Smith} used multiwavelength imaging data to study the physical properties on a sample of 12 hinge clumps, located in five interacting systems. They found star formation rates as high as 9 M$_{\odot}$ yr$^{-1}$. These high-level star formation can be explained by converging gas flows in well defined pile-up zones, which is induced by orbit crowding, as shown by the analytic models developed by \cite{2012Struck}.  Hinge clumps have central optical sources as large as 70 pc and its stellar populations span in a different range of ages, instead of a single stellar population \citep{2014Smith}. The fate of hinge clumps is unclear and it could depends on the properties of each interacting system. Therefore, a detailed analysis on these star-forming sources systems is relevant to understand its physical and kinematical properties. These reasons motivated us to perform an in-depth spectroscopic study of the most brightest star-forming region located in the compact group of galaxies NGC\,6845, by using integral field spectroscopy and archival imaging data. This region has a H$\alpha$ luminosity comparable with star-forming sources located in the Antennae merging system. Actually, it is comparable to the most luminous sources in the sample of 700 star-forming complexes belonging to interacting galaxies, as studied by \citet{Smith2016}, transforming this object in a very interesting system to understand in more detail the resolved properties of hinge clumps. 

The compact group of galaxies NGC\,6845 has been studied by \cite{1969Klemola}, \cite{1973Graham}, \cite{1979Rose}, \cite{Rodrigues99}, \cite{2003Gordon}, \cite{2015Olave-Rojas}, \cite{2021Gimeno}. It is composed by 4 interacting galaxies labeled A, B, C and, D (as it is shown in  Figure \ref{figfov}), being the main galaxy of this group NGC\,6845A,  which displays the most prominent and extended tidal tail of the system. At the outskirts of its disk, hinted in a blue square box on Figure \ref{figfov}, there is a extremely bright source 
located at the base of the northern tidal tail
, which is the most intense source of all 
\hii regions visible in the disk and tidal tails of this group. It was firstly reported by \cite{1973Graham} (knot ``a'' in their work), who described it as \emph{"one exceptional and large knot"}. The spectroscopic observations developed by these authors revealed strong emission lines. Later, using spectroscopic information, \citet{1979Rose} determined its heliocentric velocity, measuring a value of v$_{h}=6170 \pm 40$ km~s$^{-1}$. \cite{Rodrigues99} found that this region (object number 7 in their work) is young ($\sim$ 5 Myr) and bright ($\sim~10^{40} \text{erg~s$^{-1}$})$. Besides, these authors suggest that this source is associated with a star-forming complex at the end of a bar in NGC\,6845\,A. \citet{2003Gordon} studied the H~{\sc i} distribution on this system, finding that the  20cm continuum peaks at the position of this bright star-forming region. \citet{2015Olave-Rojas} used Gemini/GMOS imaging and spectroscopic data to derive the physical properties of the \hii regions in NGC\,6845, where the brightest star-forming region was named ``26''. These authors found  that this source is ionized by star-formation, having a quite high extinction and an oxygen abundance of 12+log(O/H)$\sim$8.45. Hereafter the extra-nuclear star-forming region analyzed on this report will be named ID26, following \cite{2015Olave-Rojas}. More recently, \cite{2021Gimeno} studied the compact group NGC 6845, where the authors focussed on the lenticular galaxies NGC 6845C and NGC 6845D.

The paper is structured as follows. In section \ref{data} we present the data. In Section \ref{sec:properties} we present the physical properties of the ID26. In Section \ref{sec:kinematic} we present the kinematic analysis of the ID26 region from a local and global point of view. In Section \ref{sec:discussion} we discuss the possible scenarios to explain the formation of this region. Finally, in Section \ref{sec:conclusions} we present our main conclusions. Through this paper the adopted distance to NGC\,6845 is 91.79\,Mpc \citep{Fixsen96}.

\begin{figure}
\centering
\includegraphics[width=\columnwidth]{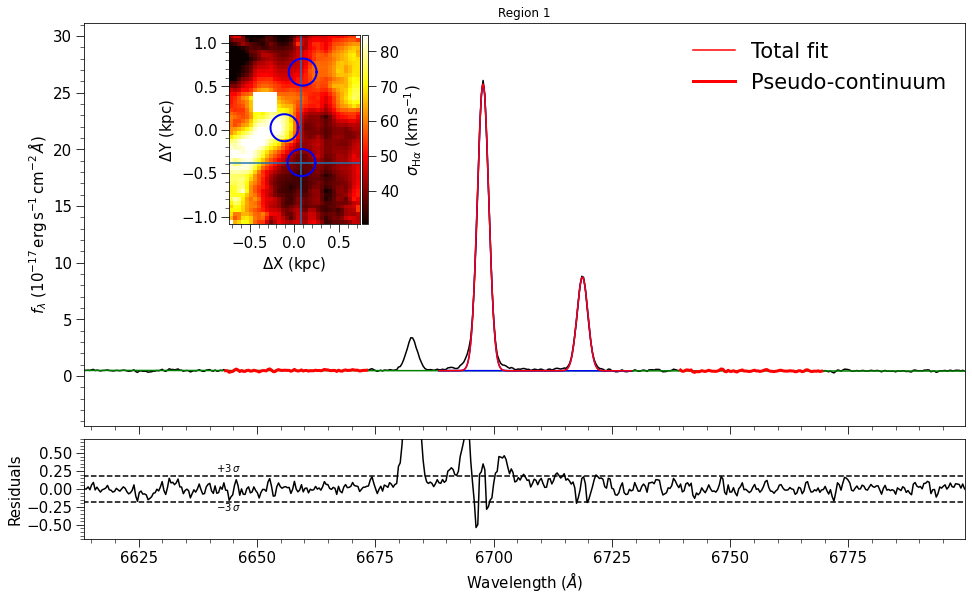}
\caption{Example of H$\alpha$ line profile for the central spaxel of Reg. 1. The main panel show the observed spectrum (black lines) within the H$\alpha$ window. The red line is the fitted Gaussian profiles for H$\alpha$ and [N{\sc ii}]\,$\lambda\,6583$ lines, and the red bold intervals are the continuum windows used to fit the continuum. There are also a bottom subpanel showing the residual of the fit. The inset show the $\sigma_{{\rm H}\alpha}$ map (in a large scale in the top-right panel of Figure \ref{fig_vel_field}) with a blue courser pointing the single spaxel where the spectrum was taken. The blue circles labeled the regions analyzed, see details in the text.}
\label{fig_line_prof}
\end{figure}

\section{Data}\label{data}

\subsection{Archival optical imaging data: Gemini and Hubble Space Telescope}

Optical imaging (\textit{u\pr}, \textit{g\pr} and \textit{r\pr}-band) of the group NGC\,6845 has been obtained at the Gemini South Observatory, using the Gemini Multi Object Spectrograph (GMOS, \citealt{2004Hook}) instrument on September 24, 2011 and September 16 and 18, 2012 (Program ID: GS-2011B-Q-36, PI: S. Torres-Flores). Details about data reduction can be found in \cite{2015Olave-Rojas}. In addition, we used archival \textit{Hubble Space Telescope} (\textit{HST}) Advanced Camera for Surveys (ACS) data (Program ID:15446, PI: J. Dalcanton). HST imaging of NGC\,6845 was observed with filter $F606W$, allowing us to disentangle the different clumps that belong to region ID26. On Figure \ref{figfov} we show a high-contrast \textit{r\pr}-band image of the system. For the central region of NGC\,6845A we show a false color image of the system, based on the \textit{u\pr}, \textit{g\pr} and \textit{r\pr}-band filters (blue, green and red colors, respectively). Black contours represent the 20cm continuum emission (taken from \citealt{2003Gordon}). In an inset we include a zoom into region ID26, where we show its resolved morphology, based on the \textit{HST} image. A central knot can be identified. The red rectangle represent the Field of View (FoV) of the Gemini/IFU observation.

\subsection{Spectroscopic data}

\subsubsection{Slit data}

\cite{2015Olave-Rojas} published the spectroscopic data of 28 star-forming regions located in the group NGC\,6845, where region ID26 was also included. These observations were taken with the GMOS instrument on Gemini South (Program ID: GS-2011B-Q-36, PI: S. Torres-Flores) and details about these observation can be found in \cite{2015Olave-Rojas}. In addition, region ID26 was observed with the Echelle double spectrograph Magellan Inamori Kyocera Echelle (MIKE, \citealt{2003Bernstein}) at the 6.5-m Magellan II (Clay) Telescope on Las Campanas Observatory during the night of September 26, 2014 (PI: S. Torres-Flores). These observations were taken under a seeing of $\sim$0.7 arcsec, using an effective slit width of 1 arcsec. Flux calibration was performed by observing the standard star Feige\,110 \citep{2001Bohlin} under the same observing conditions. In this article, MIKE observations are mainly used to compare the different flux calibrations obtained with Gemini (see section \ref{data_ifu}) and MUSE data (see section \ref{spec_muse}). MIKE observations of region ID26 and other extra-nuclear star-forming regions located in interacting systems will be discussed in a dedicated article (Firpo et al. in preparation).


\subsubsection{Gemini GMOS Integral Field Spectroscopy}
\label{data_ifu}

Given the compactness of region ID26, integral field spectroscopy (IFS) becomes in the ideal method to study the physical properties of this source. IFS observations were obtained at the Gemini South telescope using the GMOS Integral Field Unit (IFU, \citealt{2002Allington-Smith}) using the one slit mode configuration (program ID: GS-2016A-Q-46, PI: S. Torres-Flores) which has a FoV of 3.5"$\times$ 5" with a scale of pixel of 0.1", equivalent to a scale physics of 44.5\,pc according to adopted distance of NGC\,6845). In order to cover the main nebular emission-lines associated with this object (e.g. H$\beta$, [O{\sc iii}], [N{\sc ii}], H$\alpha$, [S{\sc ii}]), we have used the gratings R831 and B1200. In the case of the grating R831 we obtained three exposures of 1170 seconds each, centered at 6550\,\AA, 6600\,\AA\ and 6650\,\AA, under mean seeing conditions of 0.7 arcsec. For the grating B1200 we obtained 3 exposures of 1170 seconds, centered at 4400\,\AA, 4450\,\AA, and 4500\,\AA, under a mean seeing of 0.56 arcsec. Flats and arcs were taken right after each science exposure in order to minimize possible flexures that will affect the precision of our measurements. IFU/GMOS data was reduced using the standard {\sc iraf}\footnote{{\sc iraf} is distributed by the National Optical Astronomy Observatories, which are operate/d by the Association of Universities for Research in Astronomy, Inc., under cooperative agreement with the National Science Foundation. See http://iraf.noao.edu} Gemini Data reduction package. 
Cosmic rays were corrected using the python version of the {\sc La Cosmic} routine from \cite{vanDokkum01}. Finally, the sensitive curve of the spectra was done by using the standard star LTT\,7987. Instrumental widths were measured by fitting a Gaussian profile on the arcs lines. We estimated full width half maximum (FWHM) values of 0.7\,\AA\ and 1.5\,\AA\ for the B1200 and R834, respectively.


We have used MIKE observations to do a flux calibration of IFU GMOS observations. First, in order to perform a correct alignment between the MIKE and GMOS observations, we take as a reference the \textit{r\pr}-band image from GMOS \citep{2015Olave-Rojas}. Both observations were centered in the peak of region ID26. In a second step, we integrated the H$\alpha$ and H$\beta$ fluxes from GMOS data within an aperture of 1" to mimic the integrated echelle fluxes from the MIKE observation. Finally, from the ratio between these observation we obtained the respective calibration factors.

\subsubsection{MUSE archival complementary data}
\label{spec_muse}

In order to have a global view of the system and to clarify the origin of region ID26, we used archival data observed with the Multi Unit Spectroscopic Explorer (MUSE, \citealt{2003Henault}, \citealt{2004Bacon}, \citealt{2010Bacon}), at the Very Large Telescope (VLT) of the European Southern Observatory (ESO). Observations of NGC\,6845 were obtained in the wide field mode (Program ID 0103.A-0637(A), PI: B. Husemann), reaching a FoV of 3.34 arcmin$^2$, covering NGC\,6845~A and NGC\,6845~B. The data was acquired under a mean seeing of 1.8 arcsec, with a total exposure time of 2640~s, covering from 4750 to 9350\,\AA. We note that all the individual analysis on region ID26 will be performed by using the GMOS IFU data mainly due to the higher spatial resolution.

\section{Physical properties}\label{sec:properties}

\subsection{Emission line maps}

The emission line maps are computed by using  {\sc python} package 
 \textsc{astrocubelib}\footnote{\url{https://gitlab.com/joseaher/astrocubelib}}\citep{astrocubelib}. The line fluxes  are calculated integrating the area of line profiles within a range of 10\,\AA\, around the Gaussian peak of the line. The program also provides the Gaussian centroid, equivalent width and Gaussian $\sigma$ of each emission line. Continuum maps are computed integrating spectral windows of 30\,\AA\, distant 20\,\AA\, from the Gaussian centroid of the line. However, in the case of H$\alpha$ and [N{\sc ii}] $\lambda$6583, which were fitted simultaneously, the blueshifted spectral windows is distant of 25\,\AA\, from  H$\alpha$ Gaussian centroid to avoid the contamination of the continuum for [N{\sc ii}] $\lambda$6548 line. The continuum contribution in the emission lines then is computed fitting a linear function on the continuum windows.

In Figure \ref{fig_line_prof} we show an example of the fitting process in the case of the H$\alpha$ and [N{\sc ii}]\,$\lambda\,6583$ emission lines. We note that due to redshift of the galaxy the [S{\sc ii}] $\lambda$6731 line
falls into a sky band, then suffering an important absorption. Therefore, we do not obtain electron densities for region ID26.

\begin{figure*}
\centering
\includegraphics[width=0.27\textwidth]{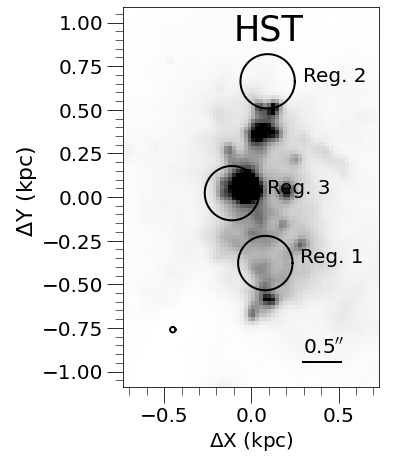}
\includegraphics[width=0.30\textwidth]{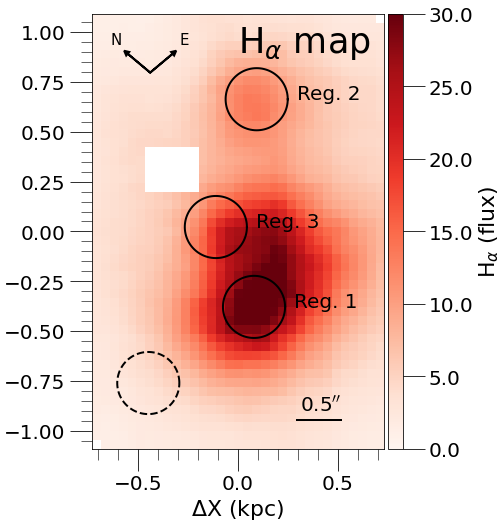}
\includegraphics[width=0.304\textwidth]{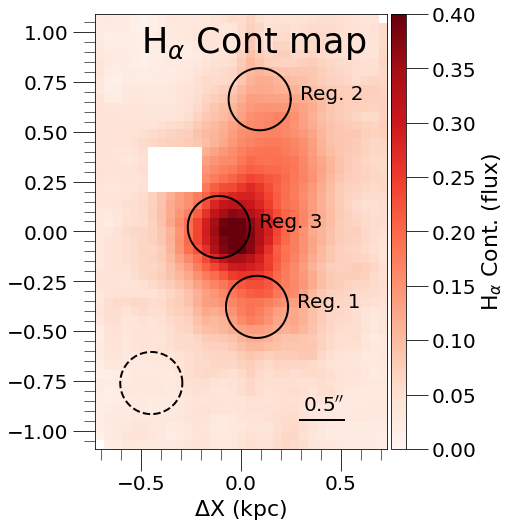}
\includegraphics[width=0.34\textwidth]{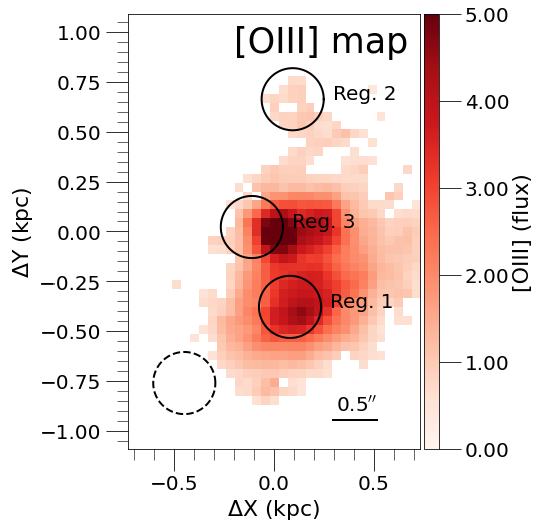}
\includegraphics[width=0.325\textwidth]{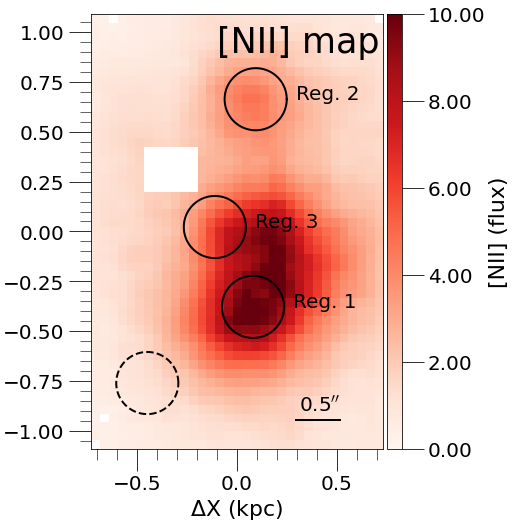}
\includegraphics[width=0.32\textwidth]{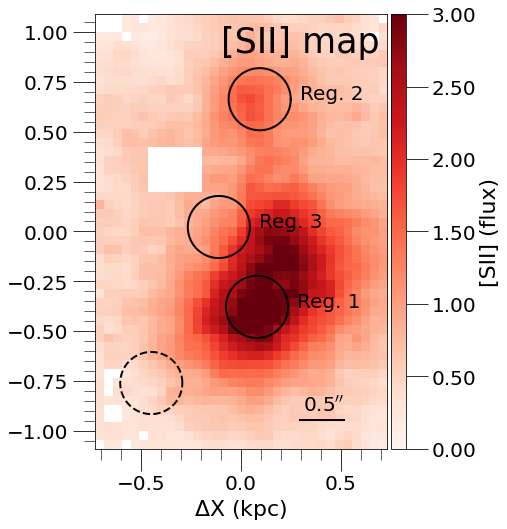}
\caption{Flux maps of region ID26. Top panels, from
left to right: 
HST
H$\alpha$ map, integrated continuum flux around H$\alpha$ line maps.
Bottom panels, from left to right: [O{\sc iii}]\,$\lambda$5007, [N{\sc ii}]\,$\lambda$6583 and [S{\sc ii}]\,$\lambda$6731 maps. The flux maps are in 
$\times10^{-17}$ erg\,s$^{-1}$\,cm$^{-2}$\,\AA$^{-1}$\,pixel$^{-1}$\,units.
The intensities in the maps are show in different ranges to show up the substructures in ID26 region. The dashed circle at right-bottom corner represents the seeing. The horizontal bar represent the angular scale. All IFU/GMOS panels show a white square which corresponds to a bad fibers in the IFU/GMOS.}
\label{fig_line_fluxes}
\end{figure*}

On Figure \ref{fig_line_fluxes} we show the HST image (top left), H$\alpha$ map (top middle), the continuum map (top right), the [O{\sc iii}] map (bottom left) and the [N{\sc ii}] and [S{\sc ii}] maps (bottom middle and bottom right, respectively). H$\alpha$ map shows a strong peak at the south direction (hereafter named Reg. 1) and a secondary knot at north-east direction (hereafter named Reg. 2), which are represented by black circles of 0.7\,arcsec in diameter in Figure \ref{fig_line_fluxes} (consistent with the seeing). The [N{\sc ii}]\,$\lambda$6583 and [S{\sc ii}]\,$\lambda$6731 emission line maps present a similar morphology. The H$\alpha$ continuum map shows a strong peak at the center of region ID26  (hereafter named Reg. 3), in agreement with the high-resolution image of the \textit{HST} (see Figure \ref{figfov}).  It is worth nothing that this central clump is made up, in reality, of many knots as seen in \textit{HST} image. On the other hand, the [O{\sc iii}] map display two peaks at the center and at west direction. This morphology resembles the structure of star-forming complexes located in interacting systems like HCG\,31 \citep{2015AlfaroCuello}, where the feedback of the continuum source seems to be triggering star formation. In the left panel of Figure \ref{fig_line_fluxes}, we presented the HST image to provide a clear overview of the spatial structure within the ID26 region. However, it is important to note that directly comparing this image with the GMOS observations can be challenging due to the difference in spatial resolution between the two images. Nonetheless, if we degrade the HST image, its morphological distribution becomes comparable to that of the continuum map obtained from the GMOS observations. The study regions are overlaid onto the HST image for reference.


\subsection{Gas extinction}

In order to correct H$\alpha$ fluxes by extinction, we have followed the same procedure of \citet{oliveira22}. We first applied a Voronoi binning on the H$\beta$ image \citep{2003Cappellari}. Our objective was to achieve a target signal-to-noise (SN) of 10 per tessellation bin. Subsequently, we generated an analogous tessellation map on the H$\alpha$ image. In the case of the theoretical H$\alpha$/H$\beta$ ratio, we assume a value of 2.86, which corresponds to the recombination case B, for an electronic temperature of 10\,000 K \citep{1989Osterbrock} and we used the extinction curve published by \citet{1989Cardelli}. 

\begin{figure*}
\centering
\includegraphics[width=0.4\textwidth]{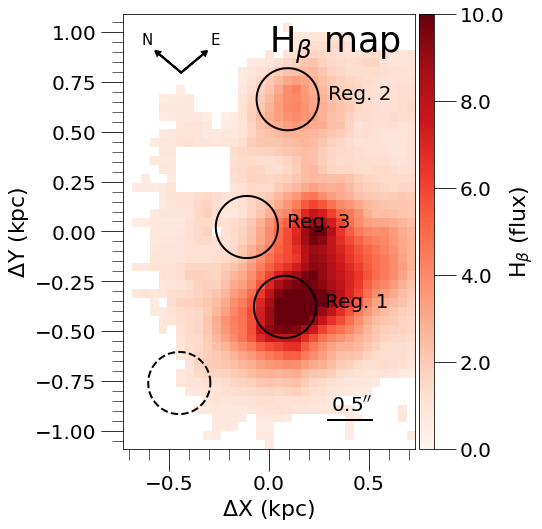}
\includegraphics[width=0.38\textwidth]{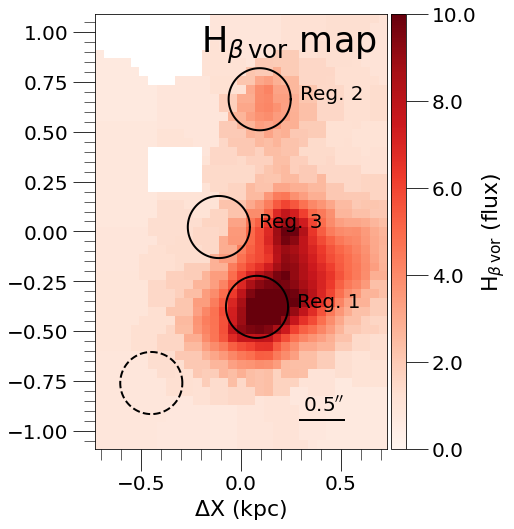}
\includegraphics[width=0.4\textwidth]{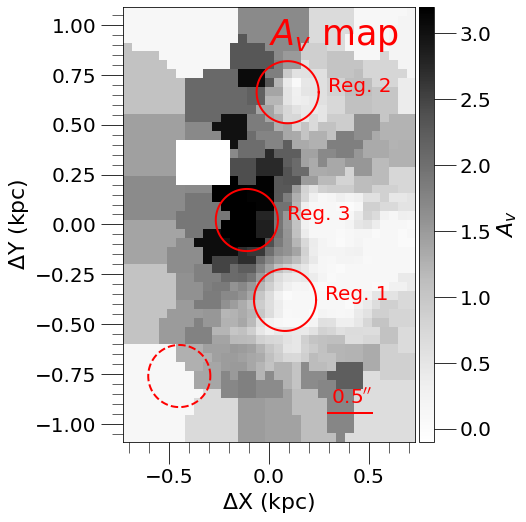}
\includegraphics[width=0.39\textwidth]{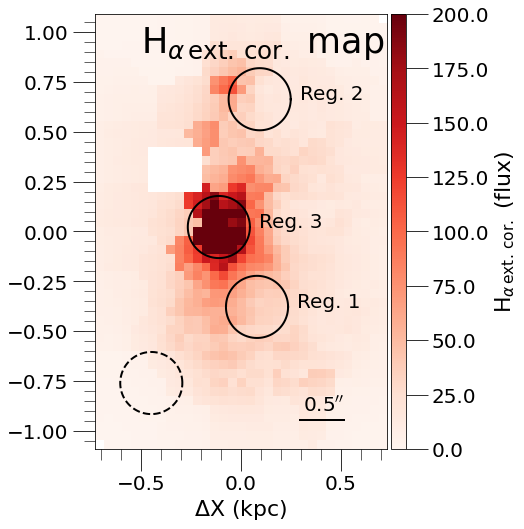}
\caption{Top left panel: H$\beta$ map. Top right panel: H$\beta$ tessellation map. Bottom left panel: extinction map in magnitudes. Top right panel: H$\alpha$ extinction corrected map. The flux maps are in $\times10^{-17}$ erg\,s$^{-1}$\,cm$^{-2}$\,\AA$^{-1}$\,pixel$^{-1}$\,unit.
The intensities in the maps are show in different ranges to show up the substructures in ID26 region. The dashed circles at right-bottom corners represent the
seeing. On each panel, the black bar show the angular scale. 
All panels show a white square which masks a set of bad fibers in the IFU/GMOS.}
\label{fig_ext}
\end{figure*}

Figure \ref{fig_ext} displays the results of the extinction correction process. As a reference, on top left panel we show the H$\beta$ map. Top right panel show the H$\beta$ map, once Voronoi tessellation was applied. Bottom left panel show the extinction map, $A_v$, in magnitudes. Bottom right panel display the H$\alpha$ map corrected by internal extinction. A comparison between the extinction map and the inset shown in Figure \ref{figfov} demonstrates the recovery of the dust lane passing through region ID26. The extinction in this particular region is around 3 $A_v$ magnitudes. Also, a comparison between the observed H$\alpha$ map (left panel in Figure \ref{fig_line_fluxes}) and the extinction corrected H$\alpha$ map shows that the most luminous clump of region ID26 is being obscured by the dust lane.

\begin{figure*}
\centering
\includegraphics[width=0.9\textwidth]{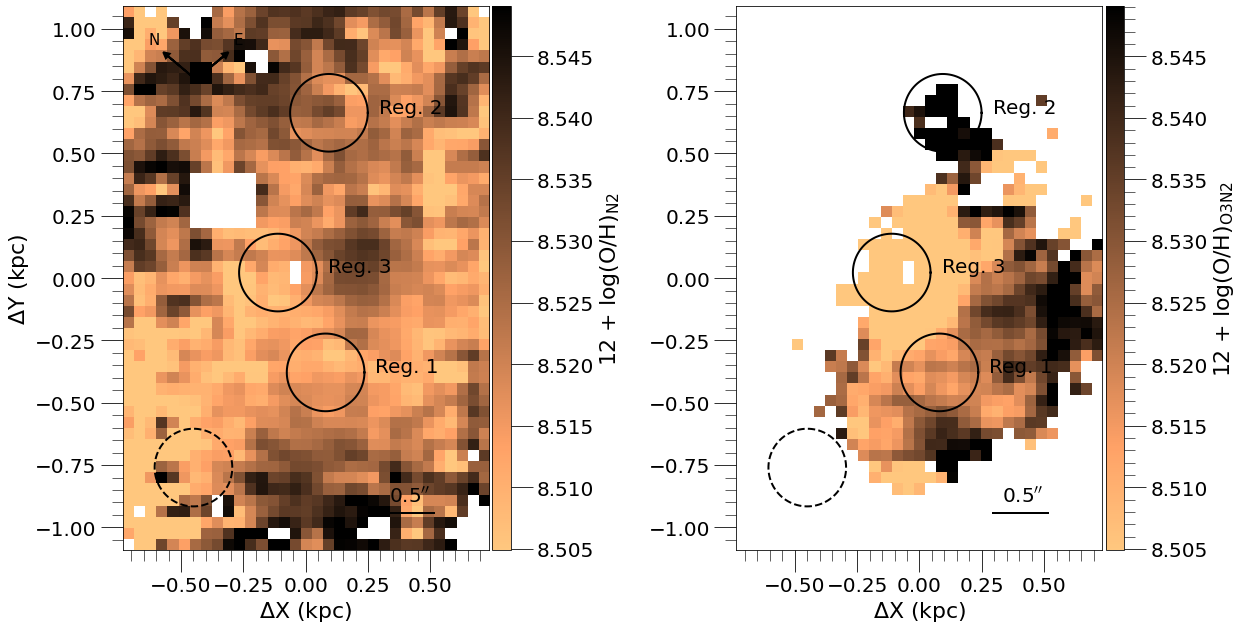}
\caption{Metallicty maps of region ID26.
Left panel: oxygen abundance map estimated from the N2 index \citep{1994StorchiBergmann}. Right panel: oxygen abundance map estimated from the O3N2 index \citep{1979Alloin}. The dashed circle at right-bottom corner represents the seeing and the bar at left-bottom corner is the angular scale. All panels show a white square which corresponds to a bad fiber in the IFU/GMOS.}
\label{fig_metallicity}
\end{figure*}

\begin{figure*}
\centering
\includegraphics[width=0.9\textwidth]{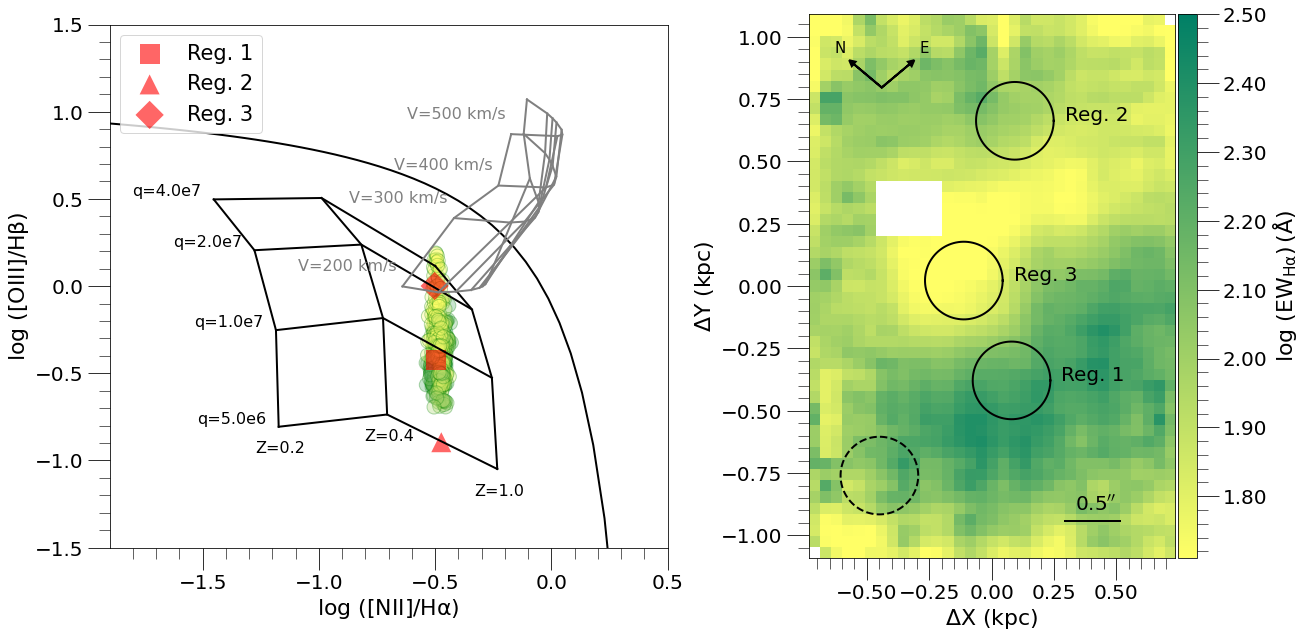}\\
\caption{Left panel: ionization mechanism diagnostic diagram in which we plot the emission line flux ratios [O{\sc iii}]/H$\beta$ $vs$  N{\sc ii}]/H$\alpha$ for all spaxels covered by IFU/GMOS FoV. The integrated points (black symbols) corresponding to the analyzed Regs. 1, 2 and 3. The spaxels points are color coded following the EW$_{{\rm H}\alpha}$ map (shown in the right panel of this figure). The black and grey grids correspond to the photoionizing and shock-ionizing models, respectively. The grid vertices of photoionizing model were compute in a range of metalicities of 0.2, 0.4, and 1\,Z$_{\odot}$, the ionizing parameter of $5\times10^6$, $1\times10^7$,
 $2\times10^7$, $4\times10^7$ \,cm\,$s^{-1}$, with a N$_\textrm{e}$ of 10\,cm$^{-3}$, and for a instantaneous burst \citep{2001Kewley}. The grid vertices of shock-ioionizing model were calculated in a range of
 shock velocities of 200, 300, 400, 500\,km\,s$^{-1}$, the 
 magnetic field from left to right of 0.0001, 0.5, 1.0, 2.0, 3.23, 4.0, 5.0, 10.0 $\mu$G, with a N$_\textrm{e}$
 of 1\,cm$^{-3}$, with a metallcity of 1\,Z$_{\odot}$, and a composition of shock+precursor regions \citep{allen08}. As reference is also display
 the maximum photoionization line (in black) from  \citet{2001Kewley}.
Right panel: the EW$_{{\rm H}\alpha}$ map.
The dashed circle at right-bottom corner represents the seeing. The bar at left-bottom corner is the angular scale. The white square corresponds to a bad fiber in the IFU/GMOS.}
\label{fig_bpts}
\end{figure*}

\begin{figure*}
\centering
\includegraphics[width=0.9\columnwidth]{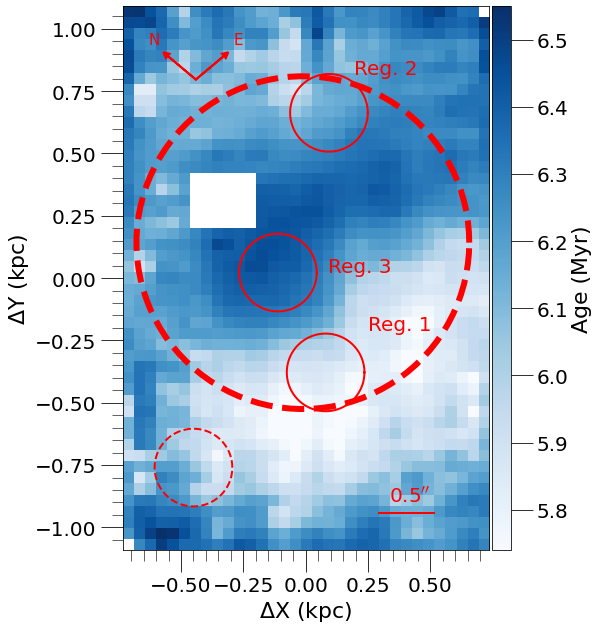}
\includegraphics[width=0.88\columnwidth]{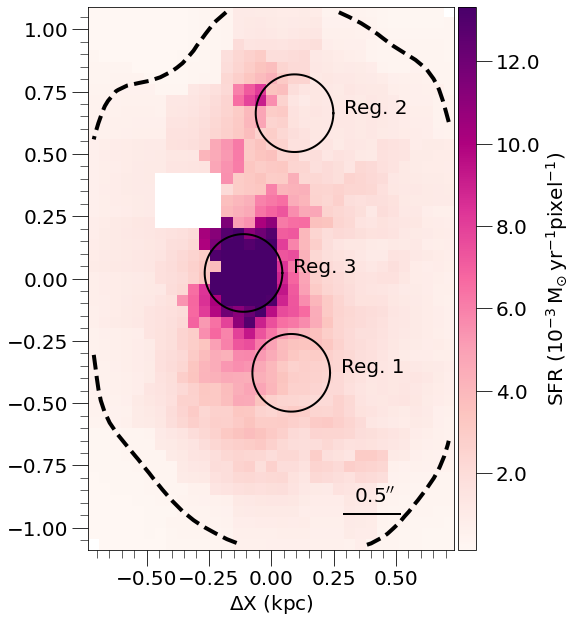}\\
\includegraphics[width=\textwidth]{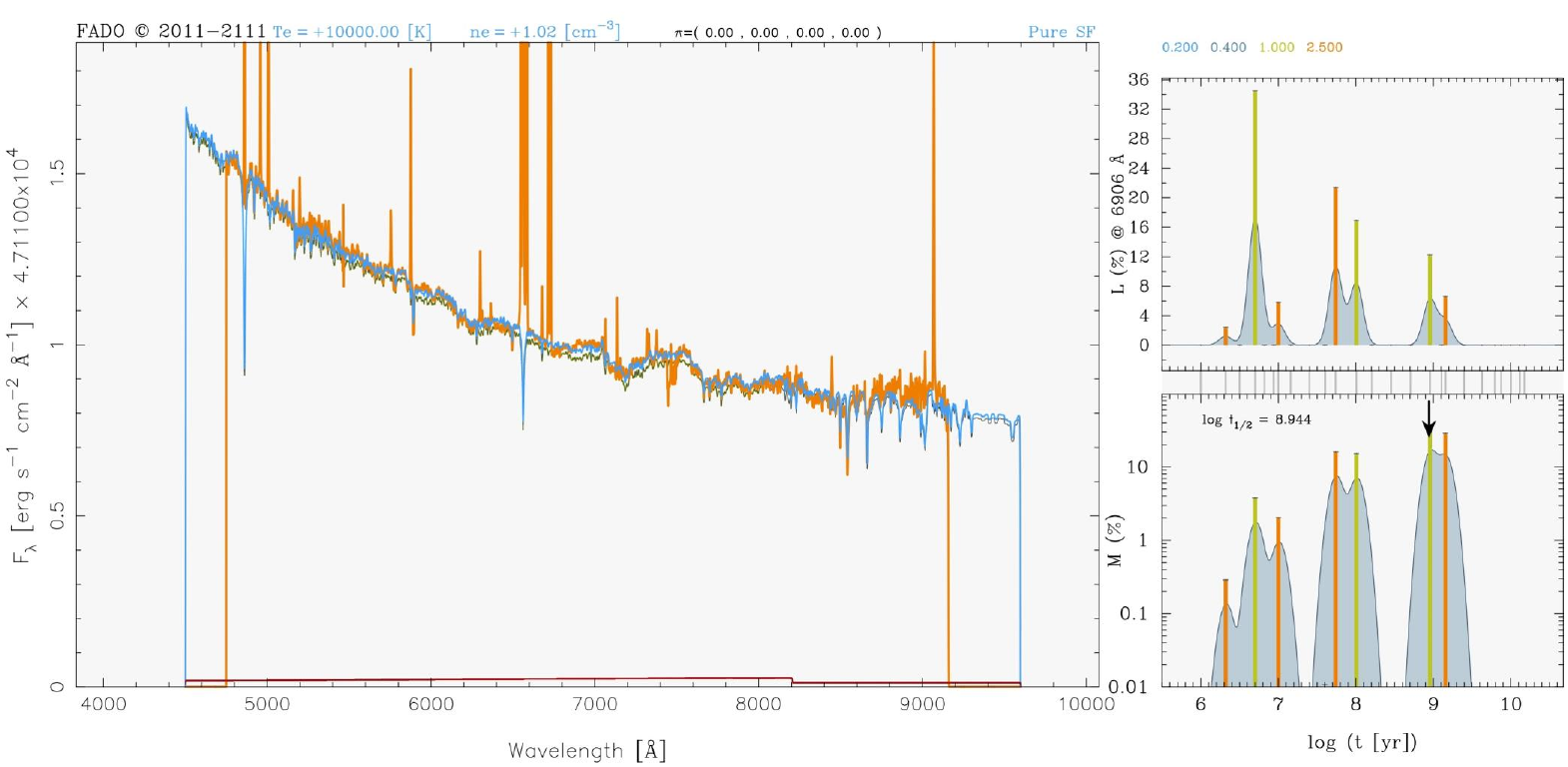}
\caption{Top Left panel: ages derived from H$\alpha$ equivalent width. The bold dashed red circle display the ring-like structure encompassing the central clump of ID26 region.
The dashed circle at right-bottom corner represents the seeing area and the bar at left-bottom corner is the angular scale. Top right panel: Star formation rates, based on H$\alpha$ luminosities. The isophote intensity, black dashed contour,
is a hundredth of the peak of H$\alpha$ map, and delimited ID26 region. The top panels show a white square which corresponds to a bad fiber in the IFU/GMOS. Bottom left panel: Standard FADO output of the MUSE integrated spectra of region ID26 (orange line). On the same spectrum we plot the best-fitting synthetic SED derived by FADO (light blue line) which is composed of stellar and nebular continuum emission (dark gray and red lines, respectively). In the upper part of that panel are the electronic temperature ($T_e$), electronic density ($n_e$) and the probability ($\pi$) of the spectrum to fall in the locus of star-forming, composite, LINER and Seyfert galaxies in the BPT diagram. Bottom upper right panel: FADO output plot showing luminosity fractions. The numbers of different colours correspond to different metallicities and the vertical bars $\pm$ 1$\sigma$. Bottom bottom right panel: stellar mass fraction of the synthetic stellar populations. The thin-gray lines connecting both diagrams correspond to the ages of synthetic stellar populations.}
\label{fig_ages_fado}
\end{figure*}

\subsection{Oxygen abundances}
\label{sec_metallicity}

Using the emission line maps previously described, we derived oxygen abundance maps by using the  N2 (\citealt{1994StorchiBergmann}) and O3N2 \citep{1979Alloin} indexes and the calibrations proposed by \cite{marino2013} which were estimated using integral field observations from the Calar Alto Legacy Integral Field Area survey (CALIFA, \citealt{2012Sanchez}). \cite{marino2013} report a dispersion of 0.18 and 0.16 dex in their calibration for the O3N2 and N2, respectively.
We do not adopt the direct method given we did not cover the [O{\sc ii}] auroral line.
In Figure \ref{fig_metallicity}  we show the oxygen abundance maps for region ID26, based on N2 and O3N2 indexes (left and right panels, respectively). 
The maps show small degrees of variations, between the percentiles 16th and 84th across the field of view, of $12+\log({\rm O/H})_{{\rm N2}}=8.52^{+0.01}_{-0.01}$ and $12+\log({\rm O/H})_{{\rm O3N2}}=8.52^{+0.02}_{-0.03}$, and both maps have  the same mean value of the metallicity. Assuming a solar metallicity (Z$_{\odot}$) of \mbox{12+log(O/H)$_{\odot}$=8.89} \cite{anders89}, the mean value of the ID26 region is 0.4\,Z\,$_{\odot}$ (based on O3N2 calibrator). Our values are consistent within 1$\sigma$ with those previously estimated by \cite{2015Olave-Rojas} for this source. 

In the case of the N2 index we do not find significant differences: 
$12+\log({\rm O/H})_{{\rm N2}}$ of 
$8.51\pm0.01$, $8.52\pm0.01$, $8.51\pm0.01$ for Reg.~1, Reg.~2 and Reg.~3, respectively. In this case, we consider as uncertainty the dispersion on the measurements inside each region. On the other hand, for the O3N2 index, the abundance derived for region Reg. 3 ($12+\log({\rm O/H})_{{\rm O3N2}}=8.43\pm0.03$) is 0.1 dex lower than values determined for Regions 1 and 2 ($12+\log({\rm O/H})_{{\rm O3N2}}$=$8.52\pm0.1$ and $8.56\pm0.1$, respectively).


 
In a large-scale context, the mean value of metallicity of region ID26 seems to follow the metallicity gradient of NGC\,6845 (\citealt{2015Olave-Rojas}), which suggest that the gas that is triggering the star-formation belongs to the disk of NGC\,6845A.

\subsection{Ionization mechanism}

A ionization mechanism diagnostic diagram (hereafter BPT, \citealt{Baldwin81}) is shown in the left panel of Figure \ref{fig_bpts}. The BPT diagram is color coded following the EW$_{{\rm H}\alpha}$ map, which is shown at the right panel of Figure \ref{fig_bpts}. On the BPT diagram we have over-plotted photoionizing (black grids) and 
shock-ionizing (gray grids) models in order to explore the dominant ionization mechanism of ID26 region. These
plots were done using the {\sc python}
package {\sc astroismlib}\footnote{https://gitlab.com/joseaher/astroismlib}  \citep{astroismlib}.
In addition of the individual spaxels, we also display the integrated information for Reg. 1, 2 and 3 (shown by a square, triangle and diamond, respectively). The photoionizing and shock-ionizing models are taken from \citet{2001Kewley}  and \citet{allen08}, respectively. Both models were computed using the gas ionization code MAPPING\,III \citep{sutherland93,sutherland13}. Most of the spaxels within the observed region are ionized by star formation. This trend is noticeable across a wide range of ionization parameters (q), falling within the interval of q$=[5\times10^6,4\times10^7]$\,cm\,$s^{-1}$. Besides, there is a discernible positive gradient in the ionization parameter as one moves towards the center of ID26 (Reg. 3), which suggests a high level of star-forming activity at that area. However, it is worth noting that the central region of ID26 also intersects with the model associated with low-velocity shocks ($200-300$\,km\,s$^{-1}$). This suggests the possibility that a dual mechanism, photoionization and shock-ionization, is acting simultaneously in this region. The sources of these shocks may include a combination of supernovae and powerful stellar winds. Indeed, as discussed in Section \ref{sec:local_kinematic}, the H$\alpha$ line profiles throughout the ID26 region exhibit signatures of multiple kinematic components, strongly indicating a complex kinematic behavior, maybe due to the presence of stellar/supernovae winds.



\subsection{Dating the hinge clump: Observing multiple stellar populations}

We use the H$\alpha$ equivalent width (right panel of Figure \ref{fig_bpts}), EW$_{{\rm H}\alpha}$, to have an estimation of the ages of the different star-forming clumps that form region ID26. Observed EW$_{{\rm H}\alpha}$ were compared with predictions derived from the {\sc starburst99, SB99} model \citep{Leitherer99}. These models determine the evolution of different physical parameters for a single stellar population. In the case of the equivalent width, EW$_{{\rm H}\alpha\,{\rm SB99}}$, we compare it with our observations, for the instantaneous case  with a solar metallicity and a \cite{1955Salpeter} Initial Mass Function (IMF). We note that SB99 models that assume a continuous SFR process display different ages evolution depending on the IMF. Then, considering this issue, and for simplicity, here we adopt an instantaneous SFR process, where there is no strong dependence on the IMF. Our results are shown in the top-left panel of Figure \ref{fig_ages_fado}. As expected from a young star-forming complex, dominated by massive star formation, most of the region has ages lower than $\sim$7 Myrs. We observe that the central clump of ID26 (Reg. 3) displays older ages, having a mean age and a standard deviation of 6.40$\pm$0.04\,Myr. This central region is encompassed by a ring-like structure (dashed red circle) that is about 0.5 Myrs younger. This structure have a radius
of about  $\sim$0.67\,kpc, and inside it, we find Reg. 1 and  Reg. 2, which exhibit mean ages and standard deviations of 5.85$\pm$0.08 and 6.18$\pm$ 0.09\,Myrs, respectively. We note that our determinations provide an estimation of the age of the last star formation episode. However, we cannot say a radial age gradient really exists. Actually, \cite{2014Smith} claim that hinge clumps have a range of stellar populations having different ages, suggesting that these objects can be long-lived structures. These authors argue that inflows should feed the star formation. Within the context of an ongoing starburst episode, it is highly plausible that the ring-like structure represents a second-generation of star formation, initiated approximately 0.5 Myr later than the central region as a direct response to the strong feedback of the central source in ID26. Besides, the BPT analysis indicates that Reg. 3 could be could potentially ionized by the influence shocks produced by stellar and supernovae winds. These powerful energetic processes can trigger new bursts of star formation. Further evidence supporting this scenario can be obtained from a detailed kinematic analysis of region ID26, which is detailed in Section \ref{sec:local_kinematic}. 


In order to determine whether region ID26 has different stellar populations, we used archival MUSE data in order to produce an integrated spectrum of this region and compare it with the FADO\footnote{http://spectralsynthesis.org/index.html} (Fitting Analysis using Differential Evolution Optimization, \citealt{2017Gomes}). We adopt this strategy instead of using the Gemini/GMOS data because this later data was observed with two different gratings (B1200 and R831), which add more complexity in the analysis of stellar populations. In the bottom panel of Figure \ref{fig_ages_fado} we display the result of this approach, where we show the FADO output. The observed spectra (orange line), the best-fitting synthetic SED (light blue line), contribution of the stellar continuum (dark gray line) and the contribution of the continuum emission spectrum (red line) are shown. As can be seen, region ID26 can be reproduced by models with a large range of ages, having models of stellar population as old as $\sim$1 Gyrs, as shown in the bottom right panels. Also, there is a contribution of a very young stellar population (a few Myrs), which is fully consistent with the EW$_{{\rm H} \alpha}$ map. All these evidences hint that the object ID26 has different stellar populations and ages. Indeed, \citet{2014Smith} indicate that hinge clumps should display a range of different stellar populations and ages, given a continuous gas flow. 
Through numerical simulations, \cite{Rodrigues99} found that NGC6845 A and B interacted ~150 Myrs ago. In this way, the interaction between galaxies could have originated the intermediate-age stellar populations shown in bottom right panels of Figure  \ref{fig_ages_fado} following with a new burst of star-formation episode giving origin to the young-age stellar populations formed in situ. Finally, the old-age stellar population shown in bottom right panels of Figure \ref{fig_ages_fado} could correspond to the old-age stellar populations inherited from the disk.

\begin{figure*}
\includegraphics[width=\columnwidth]{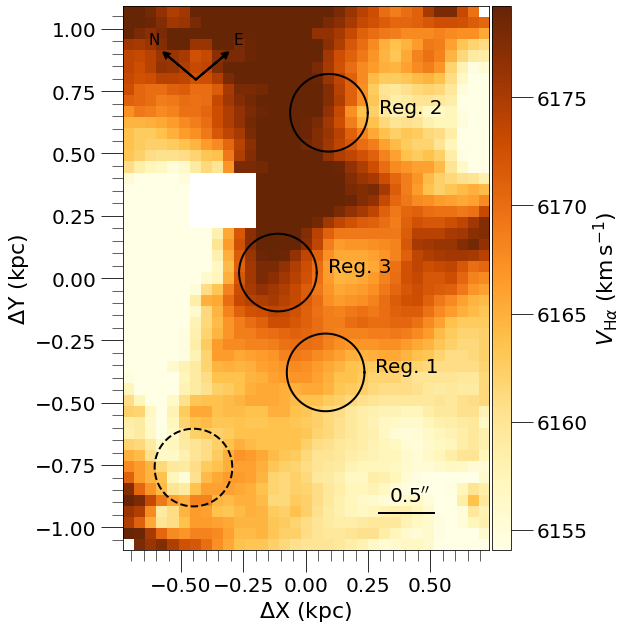}
\includegraphics[width=0.92\columnwidth]{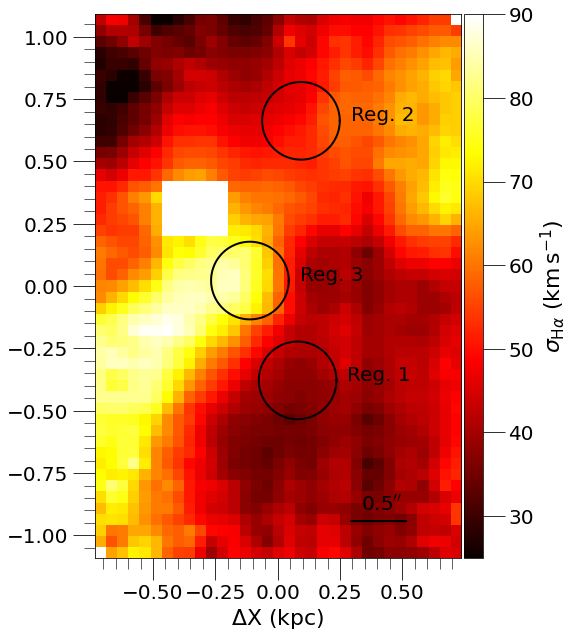}
\caption{Left panel: Velocity field near to the hinge clump. This map was derived from a single Gaussian fit on the H$\alpha$ emission line observed with Gemini/GMOS. Right panel: H$\alpha$ velocity dispersion map of the same region shown in left panel. Both top panels show a white square which corresponds to a bad fiber in the IFU/GMOS.}
\label{fig_vel_field}
\end{figure*}

\begin{figure*}
\includegraphics[width=\textwidth]{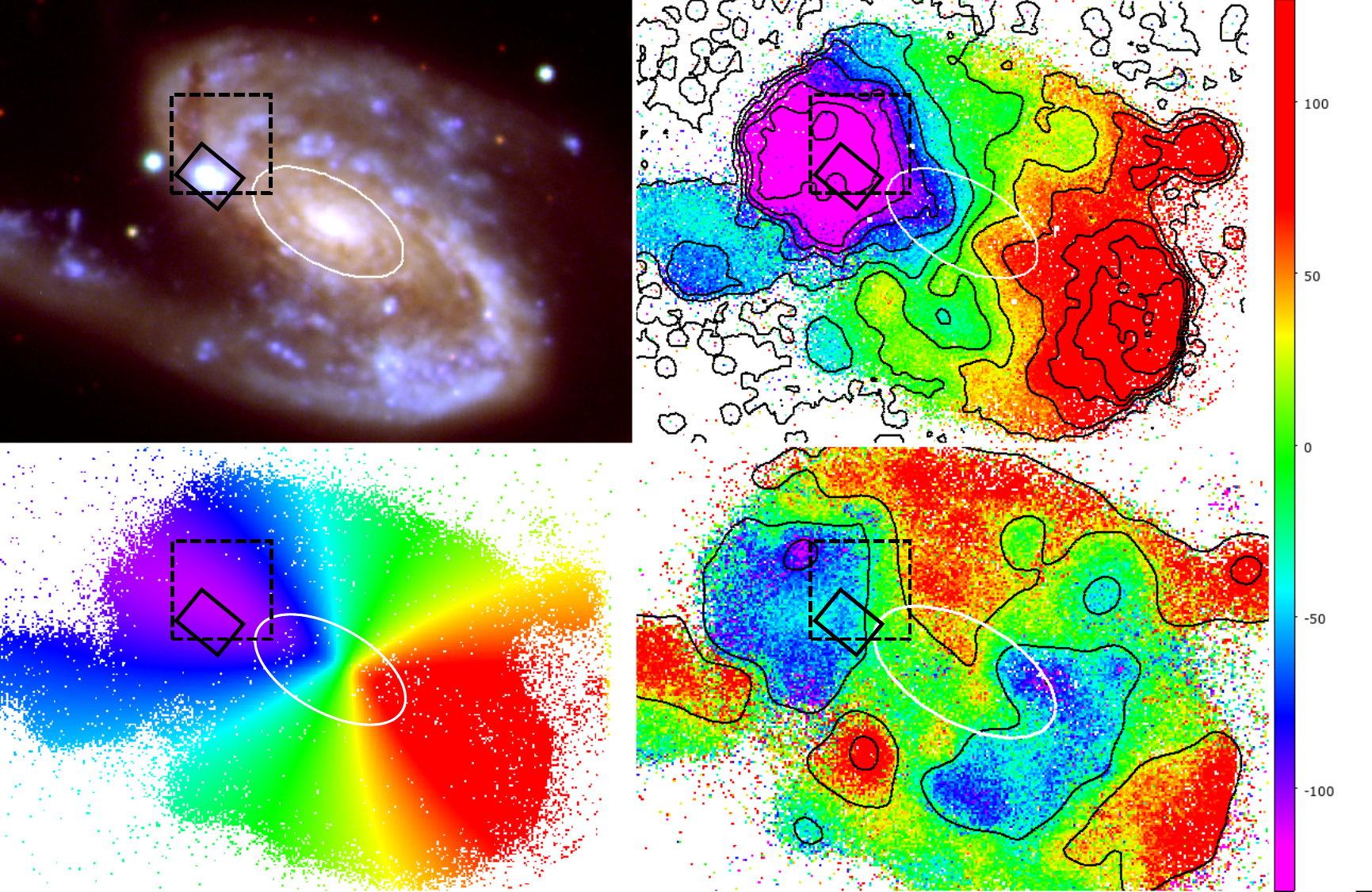}
\caption{Different maps of NGC\,6845. Top left: False color optical Gemini image. 
White ellipse represents the region that follow circular motions on the velocity field (top right panel). Top right: H$\alpha$ velocity field. Bottom left: modeled velocity field. Bottom right: Residual velocity field, derived from the observed and modeled velocity field. Bar scale represents the velocities in km s$^{-1}$ of the residual map. Black box corresponds to IFU/GMOS FoV and black dashed box corresponds to the area shows in Figure \ref{channelmap_II}.}
\label{fig_vel_field_MUSE}
\end{figure*}


\begin{figure}
\includegraphics[width=\columnwidth]{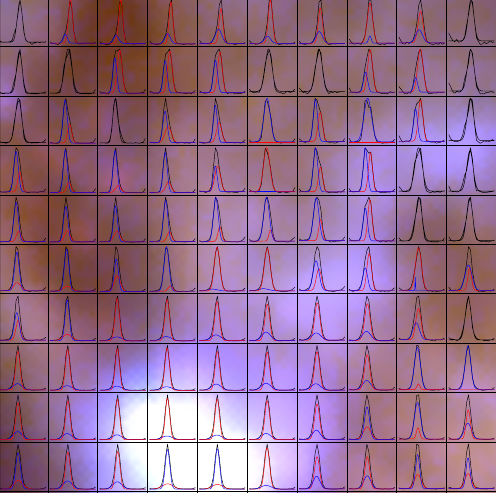}
\caption{H$\alpha$ profiles in the region of the hinge clump. Each box represents an angular size of 1 arcsec, where the emission profile was integrated on this area. Asymmetric and double profiles can be identified at the north of the hinge clump ID26. We fit two Gaussians profiles on the observed emission, which are represented by blue and red profiles on each spaxel. At the north of the hinge clump these components have a velocity difference of the order of 160 km s$^{-1}$.}
\label{channelmap_II}
\end{figure}

\subsection{A bright extra-nuclear burst of star formation}
\label{starformation}

Optical imaging revels that region ID26 is the brightest source in NGC\,6845A, even more than its central region. This vigorous luminosity is currently associated with a starburst event. On the top right panel of Figure \ref{fig_ages_fado} we show the star formation rate map for this source, which was derived from the H$\alpha$ luminosity and using equation 2 of the calibration proposed by \cite{kennicutt98}. In order to compare the SFR of region ID26 with other systems (e.g. \citealt{2014Smith}), we delimited region ID26 inside the isophote one hundredth of the peak of H$\alpha$ map. This boundary is displayed in the SFR map (black contour in Figure \ref{fig_ages_fado}) and has an equivalent radius ($R_{eq}$) of 1.9" (equivalent to 0.9\,kpc), which was estimated from the area of the isophote one hundredth of the peak of H$\alpha$ map where $R_{eq} = \sqrt{(\text{Area}/\pi)}$. We obtain a integrated SFR= 3.4\,M$_{\odot}$ yr$^{-1}$, which would placed region ID26 among the brightest hinge clumps previously detected (indeed, it is as bright as the fifth most luminous hinge clumps studied by \citet{2014Smith}, cf. its Table 10). By using the MUSE data we estimate that the hinge clump ID26 accounts for the $\sim$15\% of the total star formation in NGC 6845A. This contribution is similar to those found for the hinge clump sample of \citet{2014Smith} (cf. its Table 10). We note that most of the star formation in ID26 is localized in Reg. 3, central clump, which have almost $\sim21$\% of total SFR of ID26. Using this region, we derived a $\Sigma$ = 9.7 M$_{\odot}$ yr$^{-1}$ kpc$^{-2}$. This value should locate region ID26 in the starburst regime described in \citep{2010Daddi}, where gas inflows should provide gas to maintain the star formation.


On Figure \ref{figfov} we shown the 20cm continuum distribution on the compact group NGC\,6845 (black contours taken from \citealt{2003Gordon}). On this Figure we see that 20cm continuum peaks at the location of the hinge clump. This correlation was found in other hinge clumps. For instance, a hinge clump identified in Arp 256 is bright in 20cm continuum (\citealt{2002Chen}, \citealt{2014Smith}). In a similar way, the hinge clump named ``feature i'' in NGC 2207 (\citealt{2006Elmegreen}, \citealt{2014Smith}) is the most luminous radio continuum source on this galaxy (\citealt{1990Vila}; \citealt{2012Kaufman}). \cite{2012Kaufman} indicate that non-thermal radio emission is significant on this source. Further studies are needed in order to disentangle the origin of the radio continuum emission in region ID26, given its nature, where we can not discard the presence of supernova remnant on this source. ALMA data can be extremely useful to understand this intriguing source, in a similar way as was recently done for ``feature i'' in NGC 2207 \citep{2020Kaufman}.

\section{Kinematic of a hinge clump}\label{sec:kinematic}

\subsection{Localized kinematics}
\label{sec:local_kinematic}

In Figure \ref{fig_vel_field} we show the H$\alpha$ velocity field ($V_{{\rm H}\alpha}$) and H$\alpha$ velocity dispersion map ($\sigma_{{\rm H}\alpha}$) for region ID26 (left and right panel, respectively). Velocity dispersion map was corrected by instrumental width, thermal width \citep[9.1\,km\,s$^{-1}$,][]{odell88} and natural width \citep[3.0\,km\,s$^{-1}$,][]{clegg99}. The velocity field shows radial velocities that are consistent with the values determined by \cite{2015Olave-Rojas} for this region, in the range of 6150 km s$^{-1}$ - 6180 km s$^{-1}$. The entirety of region ID26 showcases a very perturbed velocity pattern, likely attributable to the ongoing and intense star-forming processes within,  therefore imprinting non-circular motions in the local kinematics.


The $\sigma_{H\alpha}$ map of region ID26 display values ranging from 30 km s$^{-1}$ to 90 km s$^{-1}$. In general, H{\sc ii} regions in disk galaxies display velocity dispersion of the order of 10 to 30 km s$^{-1}$ \citep{epinat10,law21}. The higher velocity dispersion in region ID26 could be attributed to the ongoing starburst process, where shocks and the feedback from massive stars contribute to driving up the velocity dispersion \citep[e.g.,][]{relano05,genzel08,rich11,rich15}. The mean values (and standard deviations) of $\sigma_{H\alpha}$ for Regs. 1, 2, and 3 are 53$\pm2$, 62$\pm3$, and 84$\pm7$ km s$^{-1}$, respectively. The central clump is located in a zone characterized by even higher $\sigma_{H\alpha}$ values exceeding 70 km s$^{-1}$ that looks like a ``finger''. This region runs through Region ID26 from west to east, effectively "piercing" the ring-like structure identified in the equivalent width map (cf. top-left panel of Figure \ref{fig_ages_fado}). Figure \ref{fig_line_prof} showcases the H$\alpha$ line profile for the central spaxel of Reg. 1. This profile exhibit a broader base that deviate from the fitted Gaussian profile. This feature serves as a signature of underlying high-velocity components throughout the entire Region ID26 \citep[e.g.,][]{relano05, firpo10, firpo11}. However,  the current GMOS spectral resolution does not allow us to perform a detailed multi-component analysis on these profiles.



\subsection{A larger context: MUSE insight}
 
Unfortunately,   the small FoV of Gemini data  hinders a detailed discussion of the origin of region ID26, such as whether it is an accretting satellite or a local hinge clump. To address this question, we conducted a  modelling of  the whole velocity field of NGC\,6845A, utilizing MUSE archival data. From this, we have derived the H$\alpha$ velocity field of NGC\,6845A, fitting a single Gaussian on the H$\alpha$ lines
over the data cube by using the \textsc{astrocubelib} code. On the top panels of Figure \ref{fig_vel_field_MUSE} we show the Gemini optical image of NGC\,6845A and its velocity field (top left and top right panels, respectively). As a reference, on the velocity field we have over-plotted its isovelocities. We observe that the central region of NGC\,6845A displays the typical rotating-like pattern of a disk galaxy, which can be seen in the isovelocities. This region is highlighted with a white ellipse, which has a semimajor axis of 8.9 arsec, equivalent to 4.3 kpc. From this radius there is a deviation from the purely circular motion, showing a vertical perturbation which seems to be associated with a warp. Indeed, a visual inspection on the Gemini image shows a warp-like structure at the southwest region of NGC\,6845A, 
region where also starts the extended southern tidal tail of NGC\,6845A.



%

In the case of the region ID26, the MUSE velocity field shows a perturbed behavior, however, this region seems to be on the plane of the disk. In order to confirm this scenario, we modeled the observed velocity field following the same methodology proposed by \citet{hernandez13,hernandez15}, which is particularly suited for interacting systems and the {\sc python}
package {\sc astromodelinglib}\footnote{https://gitlab.com/joseaher/astromodellinglib}
 \citep{astromodellinglib}.

. First, we performed a two-dimensional modeling for the light distribution of NGC\,6845A. This modeling was based on the GMOS $r$'--band image and it includes a bulge \citep{sersic68,peng10}, a bar \citep[utilizing a 2D Ferrers modified profile,][]{binney08,peng10}, and a disc \citep{freeman70,peng10}. Subsequently, we transformed the two-dimensional fitted disk profile, which represents the dominant stellar mass component of the galaxy, into its corresponding mass density profile, assuming a constant mass-to-light ratio. To derive this constant, we used the integrated color ($g-r$) up to two effective radii of NGC\,6845A and the coefficients provided in Table 7 of \citet{bell03}. The halo component was modeled using a Navarro-Frenk-White (NFW) profile \citep{1995Navarro}. Afterward, the observed velocity field is fitted letting free the halo parameters, kinematic center and position angle of line-of-nodes. The modeled velocity field is shown at the bottom left panel of \ref{fig_vel_field_MUSE}. 
Finally, modeled velocity field was subtracted from the observed one, producing a residual velocity field, which is shown at the bottom right panel of Figure \ref{fig_vel_field_MUSE}, where we include its iso-velocities. Large velocity deviations can be observed on this residual map, specially at the locations of the vertical perturbations and at the north of the hinge clump. In particular, the hinge clump displays a radial velocity of 6152 km s$^{-1}$ which is $\sim$ 30-40 km s$^{-1}$ bluer than the velocity of the modeled velocity field at that location. The velocity difference is even larger at the north of the hinge clump, reaching values of the order of $\sim$-100 km s$^{-1}$, in the residual velocity map. However, these perturbations are still compatible with the observed ones over the entire field. 

Therefore, the velocity field suggests that region ID26 follows the kinematic of the disc of NGC\,6845A, enforcing the hinge clump scenario, on which this object was formed on the disk of the main galaxy, in this case, at the base of the northern tidal tail, due to gas inflows produced by orbits crowding (see \citealt{2012Struck}). The optical images display some warp-like features on the disk of NGC\,6845A, and it displays extended tidal tails, which seems to be out of the plane of the galaxy. In addition, the H~{\sc i} channels maps displayed by \cite{2003Gordon} showed some neutral gas at the location of the hinge clump. This H~{\sc i} emission was detected starting at the same radial velocity of our optical spectra, v$_{HI}$ $\sim$ 6157 km s$^{-1}$. This H~{\sc i} detection is blueshifted with respect to the whole H~{\sc i} emission of this system, as can be seen in Figure 10 of \cite{2003Gordon}, where they display the whole velocity field for this object. Then, the neutral gas kinematics
seems to be compatible with the kinematics derived from the ionized gas.


In order to explore in more detail the kinematic of the warm gas around the hinge clump, in Figure \ref{channelmap_II} we show the H$\alpha$ profiles of the northern region of NGC\,6845A, overplotted on the optical Gemini/GMOS image. Each box has a size of 1 arsec, where the profile represents the sum of the different H$\alpha$ profiles on this box (MUSE pixel size of 0.2 arcsec pixel$^{-1}$). Figure \ref{channelmap_II} reveals a diversity on the shape the profiles. In general, the observed profiles trace the velocity field shown in Figure \ref{fig_vel_field_MUSE}. However, at the north of the hinge clump we detect asymmetric profiles, which can affect the determination of the velocity field. Actually, double H$\alpha$ profiles can be identified. Different interacting systems display double or complex H$\alpha$ profiles (e.g \citealt{2004Amram} for HCG\,31, \citealt{Repetto2010} for NGC\,5278/79), which are typically associated with gas flows and/or are produced by the geometry of the system. To determine the radial velocity of the H$\alpha$ components we perform a emission line deblending based on the H$\alpha$ emission line. 
We found a typical velocity difference of $\sim$160 km s$^{-1}$ among the two components. In Figure \ref{channelmap_II} we have highlighted, in a schematic way, the two components that can be identified in some spaxels. Blue and red Gaussian profiles represent both components. Our finding suggests that one component (redshifted) is associated with the kinematic of the disk and the other component (blueshifted) is associated with a star-formation event that is undergoing due to some extraplanar gas that is falling into the disk of NGC\,6845A. This scenario is consistent with the H~{\sc i} detection previously described. Other interacting systems that display similar features are NGC\,2207 \citep{2020Kaufman} for example. In the case of the hinge clump in NGC\,6845A, we are witnessing a strong star-formation event, in the form of a hinge clump,  which is being enhanced due to the accretion of gas. 

\begin{figure}
\includegraphics[width=\columnwidth]{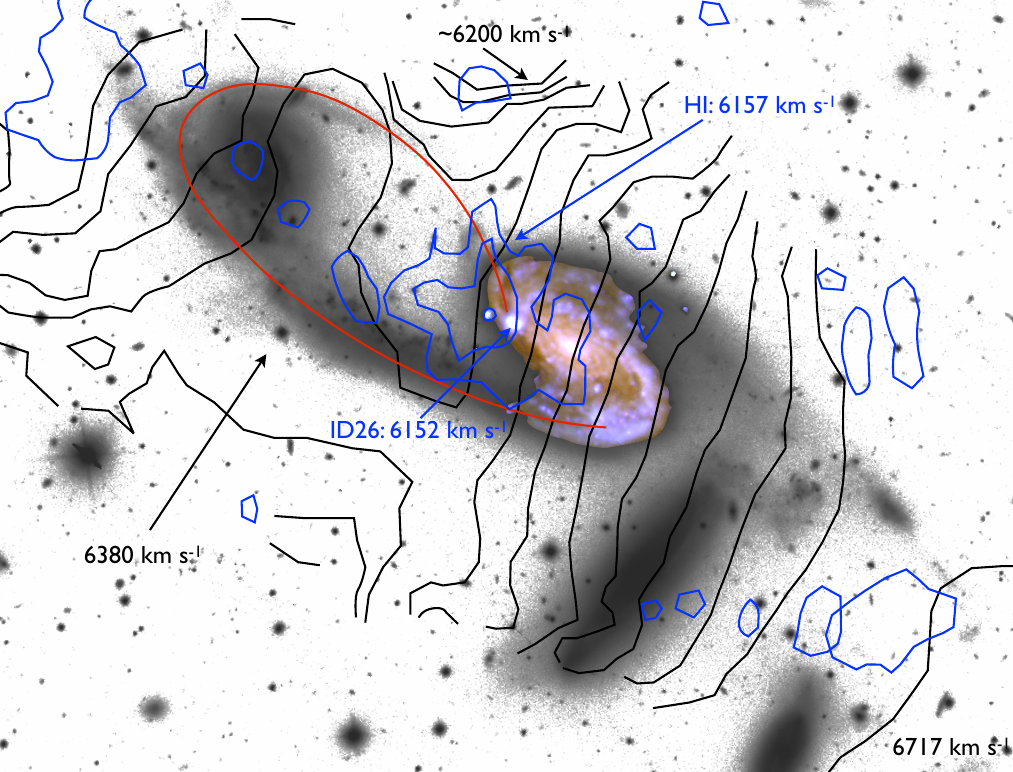}
\caption{Gemini/GMOS \textit{r\pr}-band of NGC\,6845. Black contours show the iso-velocities analysed by \citealt{2003Gordon}. Blue contours indicate the H~{\sc i} distribution at the velocity of 6152 km s$^{-1}$, as shown in the Figure 2 of \citealt{2003Gordon}. The red line shows the position of the optical tidal tail of NGC\,6845A.}
\label{info_HI}
\end{figure}

\section{Other evidences to explain the formation of the hinge clump}\label{sec:discussion}

In the case of NGC\,6845A, Figure \ref{figfov} reveals the presence of extended tidal tails. As explained above, we suggest that the hinge clump has enhanced its star formation due to some extra-planar gas flow into the disk of NGC\,6845A. In Figure \ref{info_HI} we show a high-contrast image of NGC\,6845A, where we have design a schematic representation of the eastern tidal tail of NGC\,6845A, which follow the faint stellar feature associated with the tidal tail (red line). On the same Figure we included the H~{\sc i} iso-velocities (taken from \citealt{2003Gordon}), which are represented by black lines. We also include the H~{\sc i} emission detected at 6157 km s$^{-1}$ (blue contours), as shown in Figure 2 of \cite{2003Gordon}. We include this latter H~{\sc i} emission given that it has a similar radial velocity than the hinge clump. Figure \ref{info_HI} significantly enrich the discussion to understand the origin of region ID26. The detection of some neutral gas at 6157 km s$^{-1}$ (blue contours) suggests that extra-planar gas is falling into the disk of NGC\,6845A. In addition, we can not discard that the eastern tidal tail of NGC\,6845A reach several kiloparsecs and fall down into its own disk, as represented by the red line in Figure \ref{info_HI}. This process can be producing a gas flow into the hinge clump, enhancing the star formation at this location. It is worth to note that NGC\,6845B display a radial velocity of about 6800\,km\,s$^{-1}$, therefore, the scenario proposed for the extended tidal tail can support the current observations.   

In addition, the metallicity of Region ID26 is higher than the metallicity of the regions in the tidal tails of NGC~6845A and is comparable to the metallicity of the regions in the disk of NGC~6845A (see \citealt{2015Olave-Rojas}). Moreover, \cite{2015Olave-Rojas} found that NGC 6845A displays an oxygen abundance lower than the expected for its mass and a flat metallicity gradient along its eastern tidal tail. As is shown by several authors (e.g. \citealt{2010rupke}, \citealt{2012torrey}, \citealt{2023gao}) this result could be due to the star-formation events and galactic winds triggered by the interaction/merger between galaxies. In this context, Rodrigues et al. (1999) used numerical simulations suggested that NGC 6845A had an interaction with NGC 6845B $\sim$ 150Myrs ago and according to the results found by \cite{2015Olave-Rojas}, we suggest that the interaction between theses systems produces 
an inflow of pristine gas diluting the metallicity
content in the original gas  of NGC 6845A 
flatting the metallicity gradient and enhances the star-formation activity of this system.


\section{Conclusions}\label{sec:conclusions}

On this article we report the physical and kinematical properties of a hinge clump located in the compact group of galaxies NGC\,6845. The hinge clump, named ID26, is located, on projection, at the base of a tidal tidal, in agreement with previous modeling that explain the origin of these sources \citep{2012Struck}. In addition, kinematic data allow us to rule out a possible scenario where region ID26 was a satellite galaxy. Indeed, its kinematic is consistent with the kinematic of the galactic disk of NGC\,6845A. The local metallicity abundance is also consist with metallicity gradient of the galaxy \citep{2015Olave-Rojas}. ID26 is composed by a strong H$\alpha$ emitting source, which mimic an extra-nuclear starburst. In fact, ID26 represents $\sim$15\% of total SFR of NGC\,6845. The high resolution image from HST revels that ID26 is really composed for several knots with a dominant compact source located close to the center of the H$\alpha$ emitting region. This compact source has a physical size of 70 pc, as derived from the HST image. Region ID26 has an integrated SFR=3.4 M$_{\odot}$ yr$^{-1}$, which is comparable with a few other extreme cases, e.g., sources in the Antennae galaxy \citealt{2022He} and a few other hinge clumps \citet{2014Smith}. A stellar population analysis on its integrated spectra suggests that the hinge clump is formed by multiple stellar populations instead of a single burst, having a large range of ages. This is consistent with multiple stellar populations that are seen in different star clusters. 
Besides, the analysis of the diagnostic diagram indicates that the central region can also be ionized by shocks from stellar and supernovae winds. These powerful energetic processes are able to not only ionize the surrounding environment but also act as triggers for subsequent bursts of star formation. Further evidence supporting this scenario are the high values of velocity dispersion, which  range from 30 to 90\,km\,s$^{-1}$ across this star forming region, as also found in other star-forming systems \citep[e.g.,][]{relano05,genzel08,rich11,rich15}.


\begin{acknowledgements}

We would like the thank referee Dr. R. Kotulla for his detailed revision and useful comments of this paper that greatly helped to improve its content. DEO-R acknowledges the financial support from the Chilean National Agency for Research and Development (ANID), InES-G\'{e}nero project INGE210025. JAHJ acknowledges support from FAPESP, process number 2021/08920-8. ST-F acknowledges the financial support of ULS/DIDULS through a regular project number PR222133. MDM Acknowledge to Gemini-Conicyt proyecto 3117AS0002.  Based on data obtained at the international Gemini Observatory, a program of NSF’s NOIRLab, under the scientific programs GS-2011B-Q-36 and GS-2016A-Q-46. The international Gemini Observatory at NOIRLab is managed by the Association of Universities for Research in Astronomy (AURA) under a cooperative agreement with the National Science Foundation on behalf of the Gemini partnership: the National Science Foundation (United States), the National Research Council (Canada), Agencia Nacional de Investigación y Desarrollo (Chile), Ministerio de Ciencia, Tecnología e Innovación (Argentina), Ministério da Ciência, Tecnologia, Inovações e Comunicações (Brazil), and Korea Astronomy and Space Science Institute (Republic of Korea).

{\it Software:} {\sc astrocubelib} (\url{https://gitlab.com/joseaher/astrocubelib}) \citep{astrocubelib},
{\sc astroismlib} (\url{https://gitlab.com/joseaher/astroismlib} \citep{astroismlib}), 
{\sc astromodellinglib} (\url{https://gitlab.com/joseaher/astromodellinglib} \citep{astromodellinglib}), 
{\sc astroplotlib} \citep{astroplotlib, hernandez13, hernandez15}, {\sc astropy} \citep{astropy:2013, astropy:2018, astropy:2022}, {\sc scipy} \citep{scipy}, {\sc numpy} \citep{numpy}, {\sc matplotlib} \citep{matplotlib}, and
{\sc jupyter} \citep{jupyter}.

\end{acknowledgements}

%
\bibliographystyle{aa} 
\bibliography{main} 
%






   
  



\end{document}